\documentclass[namedreferences,hyperref,optionalrh,solaromanenum]{spr-sola}

\usepackage{graphicx}                    
\usepackage{amssymb}                    
\usepackage{color}                       
\usepackage{breakurl}                         
\usepackage{booktabs}
\usepackage{natbib}
\usepackage{xcolor}

\chardef\us=`\_

\newcommand{\rev}[1]{\textbf{#1}}

\begin{document}

\title{Large-scale latitude-time relationships between the green-line corona and sunspot activity during solar cycles 18--24}


\author{Jouni J. Takalo}

\address{Taivalkuja 12 b, FI-90910 Kontio\\
\email{jojuta@gmail.com}
}
%
\runningauthor{J. J. Takalo}
\runningtitle{Relationships between the Green-line Corona and Sunspot Activity}



\begin{abstract}
We investigate the latitude--time relationship between the solar green-line corona and sunspot activity during Solar Cycles~18--24 using homogeneous coronal observations and a Gaussian representation of sunspot activity fields. The activity-field model is constructed from individual sunspot areas and latitudes in order to describe the large-scale spatial organization of solar magnetic activity more realistically than traditional sunspot-number or sunspot-area indices.
The analysis reveals a stable double-peaked latitude-dependent correlation structure centered on the active-region belts, where the strongest corona--sunspot correlations are observed. The latitude profiles remain remarkably similar from cycle to cycle, indicating that the large-scale relationship between coronal emission and photospheric activity is largely independent of cycle amplitude and parity.
The lag-correlation profiles exhibit broad positive-lag plateaus in most cycles. Although formal lag-correlation maxima occur at positive lags, surrogate-data tests indicate that these maxima are generally not statistically distinguishable from neighboring lag values. The results therefore suggest temporal persistence and memory of large-scale coronal magnetic structures rather than a precisely defined physical delay.
The Gaussian activity-field representation produces substantially stronger and more coherent correlations with the green-line corona than conventional sunspot measures, supporting the interpretation that the large-scale distribution of magnetic activity, rather than individual sunspots alone, governs the evolution of the large-scale corona.
\end{abstract}


\maketitle

\section{Introduction}

The solar green-line corona observed at 530.3~nm (Fe~XIV) is one of the most important tracers of large-scale coronal magnetic activity. The emission originates from highly ionized iron at temperatures near $2\times10^6$~K and provides an observational link between photospheric magnetic activity and the extended solar corona \citep{Rusin2002}. Since systematic coronagraphic observations began in the mid-twentieth century, the green-line corona has become one of the longest and most homogeneous observational records of the large-scale structure and long-term evolution of the solar corona.

The Homogeneous Coronal Data Set (HCDS) and the coronal index (CI) have been widely used to investigate the temporal and spatial evolution of solar activity over multiple solar cycles and remain among the longest continuous records of global coronal behavior available for solar-cycle studies. Long-term properties of the HCDS database and the coronal index were described by \citet{Minarovjech2011}, who demonstrated that the green-line corona provides a stable tracer of global solar activity. Earlier studies showed that the green-line brightness is closely related to the evolution of sunspot activity and large-scale magnetic fields \citep{Rybansky2001,Rusin2002,Rybansky2005}.

The latitude distribution of the green-line emission has attracted particular interest because coronal structures extend over a much broader latitude range than sunspots and reflect both local active-region fields and global-scale magnetic organization. Studies comparing coronal brightness with photospheric magnetic fields and sunspot distributions have demonstrated significant latitude-dependent relationships between these quantities \citep{Vernova2016,Takalo2022}. Investigations of north--south asymmetries have shown that
green-line brightness, sunspot number, and sunspot area exhibit remarkably similar temporal behavior on both long and
short time scales. Furthermore, quasi-biennial oscillations appear coherently in all these indices, suggesting that coronal
and photospheric manifestations of solar activity are governed by common large-scale magnetic processes
\citep{Badalyan2005,Badalyan2008}.

Using the HCDS database, \citet{Takalo2022} investigated the latitude--time structure of the green-line corona during Solar Cycles~18--24 and showed that the strongest coronal emission is concentrated near the active-region belts, while weaker structures extend toward considerably higher latitudes. The study also identified both equatorward and poleward migrations of coronal structures during the solar cycle. These findings support the view that the green-line corona is not merely a passive consequence of sunspot activity but represents a large-scale manifestation of the evolving solar magnetic field.

The temporal evolution of coronal activity is known to exhibit complex behavior. Although the coronal index generally follows the solar cycle, differences emerge during cycle maxima and declining phases when large-scale magnetic structures and coronal holes become increasingly important \citep{Hathaway2015,Kane2015}. Evidence for the Gnevyshev gap (GG), a temporary reduction of activity near solar maximum, has been reported in several solar and heliospheric parameters \citep{Gnevyshev1977,Feminella1997,Ahluwalia2004,TakaloMursula2020}. In the green-line corona, however, the occurrence and characteristics of the GG remain less well understood. \citet{Takalo2022} reported statistically significant reductions of coronal activity near the maxima of several cycles, suggesting that the phenomenon may also be present in coronal emission.

Recent studies have further examined temporal and spatial relationships between green-line emission and other indicators of solar activity. Latitude dependent correlations between coronal emission, photospheric magnetic fields, and solar activity indices have been reported over multiple solar cycles \citep{Vernova2016,Badalyan2014}. Long-term analyses have shown that the coronal index remains closely connected
to the global evolution of solar magnetism while exhibiting cycle-dependent behavior distinct from traditional photospheric indices \citep{Kane2015}. Furthermore, recent investigations have identified latitude-dependent correlations and temporal offsets between green-line brightness, sunspot activity, flare occurrence, and radio flux, suggesting that coronal emission behaves as a temporally persistent manifestation of large-scale solar magnetic fields rather than as an instantaneous response to photospheric activity \citep{Morgan2017,Oloketuyi2024,Oloketuyi2025}. These findings indicate that coronal emission contains information on the organization and evolution of the global magnetic field that is not fully captured by traditional sunspot indices alone.

Despite these earlier investigations, quantitative comparisons between the latitude--time structure of the green-line corona and spatially distributed representations of sunspot activity remain limited. Most previous studies have relied on global activity indices, sunspot numbers, sunspot areas, or magnetic-field distributions. Because the corona is fundamentally a large-scale magnetic structure, localized measures of sunspot activity may not optimally represent the spatial extent and temporal persistence of the underlying magnetic field. This motivates the use of continuous latitude--time activity fields when investigating the relationship between photospheric activity and coronal evolution.

In this work we construct a latitude--time Gaussian representation of sunspot activity and compare it with the green-line corona during Solar Cycles~18--24. The objectives of this study are:

\begin{enumerate}
\item to quantify the latitude dependence of the corona--sunspot relationship,
\item to investigate possible temporal delays between photospheric and coronal activity,
\item to examine cycle-to-cycle variability between Solar Cycles~18--24, and
\item to explore whether temporary reductions of coronal intensity near solar maximum are consistent with the Gnevyshev-gap phenomenon.
\end{enumerate}

\section{Data}

The homogeneous coronal data set (HCDS) is the irradiance of the Sun as a star in the
coronal green line (Fe XIV, 530.3 nm). It is derived from ground-based observations of the green corona made by the network of coronal stations (Kislovodsk, Lomnick\'{y} \v{S}t\'{i}t, Norikura, and Sacramento Peak). These indices are not, however, measured anymore in the traditional way as was made earlier at Lomnick\'{y} \v{S}t\'{i}t Observatory (former Lomnick\'{y} \v{S}t\'{i}t coronal station). The coronal intensities have been measured at 72 points at 5 degree separation starting from north pole counterclockwise around the Sun at height around 50 arcsec. The values are calibrated to the center of the solar disk to get absolute values of intensity, i.e. absolute coronal units (ACU). One ACU represents the intensity of the continuous spectrum of the center of the solar disk in the width of one {\AA}ngstr{\"o}m at the same wavelength as the observed coronal spectral line (1ACU = 3.89 Wm$^{-2}$ sr$^{-1}$ at 530.3 nm). 
Recently the corona indices were corrected mainly for the pre-1966 era. In this research we use the new reconstructed corona indices, which  were obtained from the NOAA/NGDC solar corona archive (\url{https://www.ngdc.noaa.gov/stp/solar/corona.html}). \citep{Rybansky2005}. The main period in this study is Solar Cycles 18\,--\,23.  In some cases we also use data for Solar Cycle 24. These were obtained from the Slovak Central Observatory Hurbanovo archive \citep{Lukac_2010}.

The approximately 11-year solar cycle is characterized by systematic variations in sunspot number, sunspot area, magnetic activity, and numerous other indicators of solar activity \citep{Hathaway2015}. Sunspot observations were obtained from the RGO--USAF/NOAA sunspot database available at \url{https://solarscience.msfc.nasa.gov/greenwch.shtml}.

Separate datasets were analyzed for Solar Cycles 18--24. Cycles 18--23 are complete, whereas Cycle 24 covers only the interval from January 2009 to September 2016 because later observations were not available in the adopted dataset. The structure of each sunspot record is [decimal year, sunspot area, heliographic latitude].

\section{Methods}

\subsection{Construction of coronal latitude--time maps}

The green-line coronal data were obtained from the homogeneous coronal data set (HCDS), which provides daily Fe~XIV 530.3~nm coronal intensities measured around the solar limb at 72 position angles separated by 5$^\circ$. To construct latitude--time coronal maps, the limb position angles were transformed into approximate heliographic latitudes using

\begin{equation}
\lambda = 90^\circ \cos(\phi),
\end{equation}

where $\phi$ is the position angle around the solar limb.

Since identical heliographic latitudes occur on both the east and west limbs, duplicate latitude values were combined by averaging the corresponding coronal intensities. The resulting latitude profiles were interpolated onto a regular latitude grid spanning

\[
-90^\circ \le \lambda \le 90^\circ
\]

with 2$^\circ$ spacing.

The interpolation to a 2$^\circ$ latitude grid was performed to obtain a regular latitude--time representation compatible with the Gaussian activity fields. The interpolation does not increase the intrinsic spatial resolution of the original coronal observations, which remains limited by the 5$^\circ$ sampling of the position-angle measurements. To suppress small-scale fluctuations while preserving the large-scale butterfly structure, the latitude--time maps were subsequently smoothed using moving averages in both latitude and time directions.

\subsection{Gaussian sunspot activity field}

To represent the large-scale spatial organization of solar magnetic activity, a continuous latitude--time activity field was constructed from individual sunspot observations using Gaussian kernels. Unlike traditional sunspot-number or spot-area indices, this representation preserves both the spatial distribution and temporal evolution of active regions, allowing a direct comparison with the extended coronal emission field.

The activity field was defined as

\begin{equation}
S(\lambda,t)
=
\sum_i
w_i
\exp
\left[
-\frac{(\lambda-\lambda_i)^2}
{2\sigma_\lambda^2}
\right]
\exp
\left[
-\frac{(t-t_i)^2}
{2\sigma_t^2}
\right],
\end{equation}

where $\lambda_i$ and $t_i$ denote the latitude and occurrence time of sunspot $i$, while $\sigma_\lambda$ and $\sigma_t$ represent the latitude and temporal smoothing scales, respectively. The weighting factor $w_i$ describes the contribution of each sunspot to the distributed activity field. Equation (2) defines the activity field used in this study. It is constructed by representing each sunspot group as a two-dimensional Gaussian kernel in latitude and time, analogous to Gaussian kernel density estimation \citep{Silverman1986}. The activity field is then obtained as the superposition of all kernels. This kernel-based representation provides a continuous approximation of the otherwise discrete sunspot distribution while preserving the spatial and temporal localization of solar activity.

\begin{figure}[ht]
\centering
\includegraphics[width=0.85\textwidth]{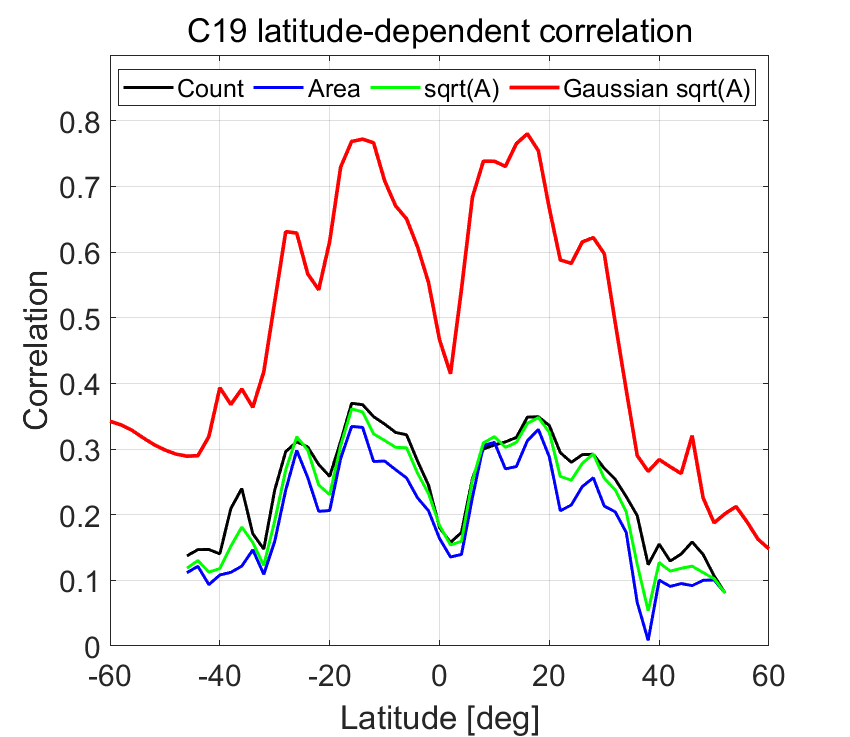}
\caption{
Latitude-dependent correlation between the green-line corona and four different sunspot activity representations for Solar Cycle 19. The black, blue, and green curves correspond to correlations obtained using sunspot counts, sunspot areas, and square-root area weighting, respectively. The red curve shows the correlation derived from the Gaussian sunspot activity field, constructed by applying spatial and temporal Gaussian kernels to the square-root weighted sunspot areas. The Gaussian representation produces substantially stronger and more coherent correlations over the active-latitude belts, demonstrating that a distributed activity field provides a more realistic description of the large-scale coronal response than localized sunspot occurrence alone.}
\label{fig:area_count_gaussian}
\end{figure}

Several weighting schemes were tested, including uniform weighting ($w_i=1$), area weighting ($w_i=A_i$), and square-root area weighting ($w_i=\sqrt{A_i}$), where $A_i$ is the sunspot area.  Figure~\ref{fig:area_count_gaussian} shows that the Gaussian activity-field representation produces substantially higher latitude-dependent correlations than the count-, area-, or square-root area representations, which motivated its adoption in the subsequent analyses.The resulting correlation structures were qualitatively similar for all weighting schemes. The square-root weighting was adopted because it preserves the contribution of larger active regions while reducing the excessive dominance of the very largest sunspots. This produces a more balanced representation of the distributed magnetic activity field and improves the stability of the latitude--time correlations between different solar cycles.

The Gaussian representation produced systematically stronger and spatially smoother correlations with the green-line corona than simple spot counts or spot areas alone. The smoothing procedure should not be interpreted as a physical diffusion model, but rather as an empirical representation of the spatially extended magnetic influence of active regions on the large-scale corona.

Unless otherwise noted, the adopted smoothing parameters were

\[
\sigma_\lambda = 2^\circ
\]

and

\[
\sigma_t = 11 \ {\rm days}.
\]

To assess the sensitivity of the results to the temporal smoothing scale, additional analyses were performed using $\sigma_t$ values of 5, 20, and 27 days. The overall correlation structure, latitude dependence, and lag-correlation profiles remained qualitatively similar for all tested values, indicating that the results are robust with respect to the choice of temporal smoothing scale. The adopted value $\sigma_t$ =11 days corresponds to a full width at half maximum (FWHM) of approximately 26 days, which is close to the synodic solar rotation period. The temporal smoothing therefore distributes the contribution of individual sunspot groups over a timescale comparable to one solar rotation. This choice suppresses short-lived fluctuations while preserving the large-scale latitude--time evolution of solar activity.
\subsection{Latitude-dependent correlation analysis}

Latitude-dependent correlations between the coronal emission field and the field of sunspot activity  were computed separately for each latitude as

\begin{equation}
R(\lambda)
=
{\rm corr}
\left[
C(\lambda,t),
S(\lambda,t)
\right],
\end{equation}

where $C(\lambda,t)$ is the coronal intensity field and $S(\lambda,t)$ is the Gaussian sunspot activity field. This procedure measures how closely the temporal evolution of coronal emission follows the corresponding latitude-dependent evolution of photospheric activity.

\begin{figure}[ht]
\centering
\includegraphics[width=1.0\textwidth]{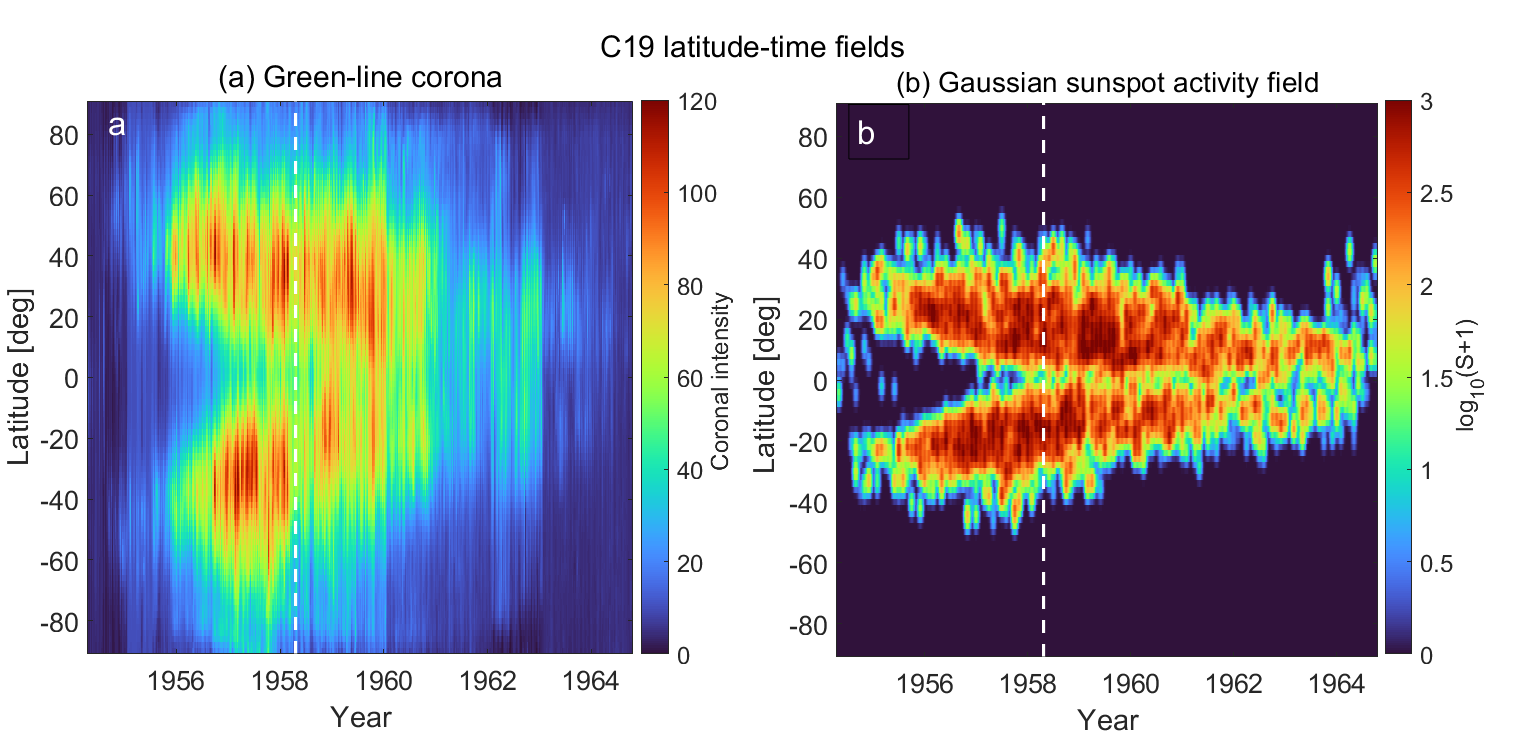}
\caption{
Latitude--time fields for Solar Cycle 19.
(a) Smoothed green-line coronal intensity field derived from daily observations of the Homogeneous Coronal Data Set (HCDS).
(b) Gaussian sunspot activity field constructed from individual sunspot areas and latitudes using spatial and temporal Gaussian kernels.
The vertical dashed line marks the approximate epoch of solar maximum. Both fields exhibit a similar butterfly-like equatorward migration pattern associated with the evolution of active regions throughout the solar cycle. The coronal emission is spatially broader and more diffuse than the sunspot activity field, reflecting the large-scale magnetic connectivity and extended structure of the solar corona.}
\label{fig:2D_figures}
\end{figure}

,
\subsection{Time-lag correlation analysis}

To investigate temporal delays between photospheric activity and coronal response, lag correlations were calculated as

\begin{equation}
R(\tau)
{\rm corr}
\left[
C(\lambda,t+\tau),
S(\lambda,t)
\right],
\end{equation}

where $\tau$ denotes the temporal lag in days. Positive lag values correspond to situations where the coronal activity follows sunspot activity with a delay. The correlations were computed over the full latitude--time fields.

To assess whether the observed lag maxima represent statistically significant delays rather than broad correlation plateaus, the difference

\begin{equation}
\Delta R = R(\tau_{\max}) - R(0)
\end{equation}

between the maximum lag correlation and the zero-lag correlation was calculated. This quantity measures the improvement in correlation obtained by allowing a temporal lag relative to the zero-lag case. The 95\%-of-maximum lag interval was defined as the range of lag values for which the lag correlation remained at least 95\% of its maximum value, that is, $R(\tau) \geq 0.95 \times R(0)$ The lower and upper bounds of this range were reported as the corresponding lag limits. These intervals describe the width of the lag-correlation plateau and should not be interpreted as conventional statistical confidence intervals.

The statistical significance of the observed lag maxima was evaluated using surrogate data. For each solar cycle, 1000 surrogate activity fields were generated by circularly shifting the complete Gaussian sunspot activity field in time relative to the coronal field by a randomly selected offset. The random shift was restricted to the interval from 300 days to $N-300$ days, where $N$ is the duration of the analyzed solar cycle, in order to disrupt the original temporal alignment while preserving the latitude structure and temporal persistence of the activity field. The lag-correlation analysis was repeated for each surrogate realization, and the statistic

\begin{equation}
\Delta R = R(\tau_{\max}) - R(0),,
\end{equation}

was calculated for the observed data and for each surrogate realization. The empirical $p$-value was then computed as the fraction of surrogate realizations for which

\begin{equation}
\Delta R_{\rm sur} \geq \Delta R_{\rm obs}.
\end{equation}

\section{Results}

\subsection{Latitude--time structure of the green-line corona and sunspot activity}

Figure \ref{fig:2D_figures} compares the latitude--time evolution of the green-line coronal intensity with the Gaussian sunspot activity field for solar cycle~19. The coronal emission exhibits broad enhancements centered at active latitudes in both hemispheres, while the Gaussian sunspot field reproduces the well-known butterfly pattern of sunspot emergence. Despite the very different appearance of the two fields, the large-scale latitude--time evolution is remarkably similar.

The strongest coronal enhancements occur at latitudes between approximate values $10^\circ$ and $30^\circ$, corresponding to the main sunspot activity belts. However, the coronal structures are spatially broader and temporally smoother than the corresponding sunspot distributions. Significant coronal emission is also present outside the main activity belts, especially at higher latitudes.

The comparison demonstrates that the Gaussian activity field captures the large-scale morphology of the coronal intensity distribution substantially better than the simpler count- or area-based representations.


\subsection{Latitude-dependent correlations}

\begin{figure}[ht]
\centering
\includegraphics[width=0.8\textwidth]{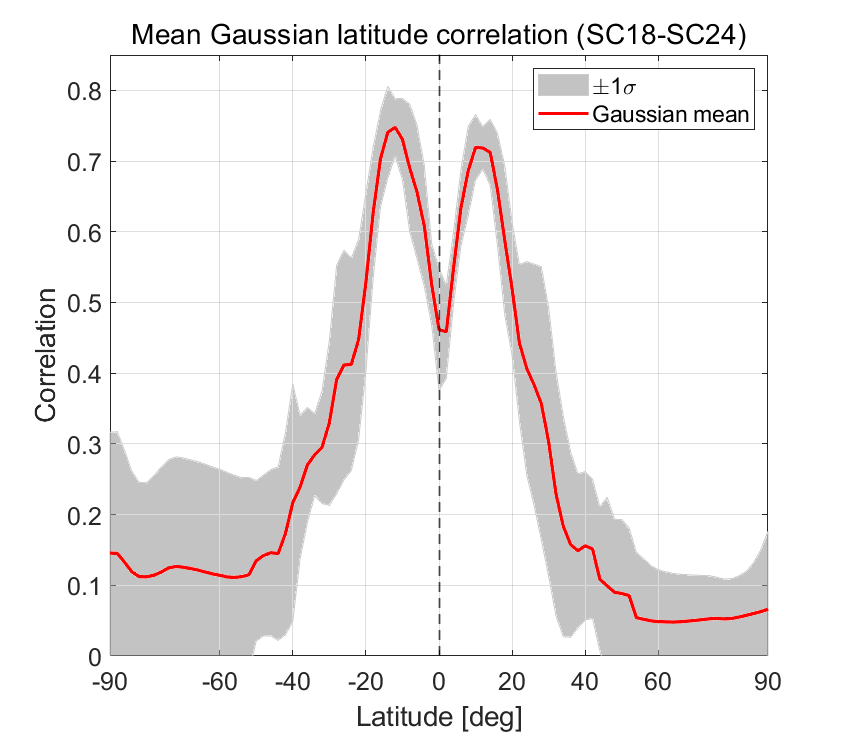}
\caption{
Universal latitude-dependent correlation profile between the green-line corona and the Gaussian sunspot activity field for Solar Cycles 18--24.
The red curve shows the mean correlation profile averaged over all analyzed cycles, while the gray shaded region indicates the corresponding $\pm1\sigma$ cycle-to-cycle variability. Correlations were calculated without temporal lag between the coronal and sunspot activity fields.
The profile exhibits a stable double-peaked structure with maxima located within the active-latitude belts near $\pm10^\circ$--$20^\circ$. Correlations decrease toward both the equator and higher latitudes, indicating that the strongest coupling between coronal emission and photospheric activity occurs within the core active-region belts. The relatively small cycle-to-cycle variability demonstrates that this latitude-dependent structure is a persistent feature of Solar Cycles 18--24.}
\label{fig:mean_correlation}
\end{figure}

The latitude-dependent zero-lag Gaussian correlation between the green-line corona and the sunspot activity field is shown in Figure~\ref{fig:mean_correlation}. The universal latitude-correlation profile averaged over cycles~18--24 exhibits two pronounced maxima near $\pm10^\circ$--$20^\circ$, corresponding to the mean active-latitude belts. Correlation decreases toward the equator and toward high latitudes, producing a characteristic double-peaked structure.

A local minimum is consistently present near the equator. This behavior likely reflects the reduced concentration of sunspots near $0^\circ$ latitude, while the corona remains relatively strong there due to the large-scale connectivity of coronal magnetic structures.

The cycle-to-cycle variability remains relatively small within the active latitude belts, as indicated by the $\pm1\sigma$ region in Figure~\ref{fig:mean_correlation}. This suggests that the latitude dependence of the corona--sunspot relationship is a stable property of solar cycles.

The Gaussian weighting method yields substantially higher correlations than either spot counts or spot areas alone. This indicates that spatial smoothing in both latitude and time provides a more realistic representation of the large-scale magnetic structures influencing the green-line corona.

The latitude dependence of the corona--sunspot correlation is broadly consistent with the latitude distribution of strong photospheric magnetic
flux reported by Vernova et al. (2016). Both quantities exhibit maxima within the active-latitude belts near $\pm15^\circ$--$\pm20^\circ$,
although the correlation profile is somewhat narrower than the magnetic-flux distribution. This suggests that the strongest corona--sunspot coupling
occurs within the core active-region belts rather than across the full latitude range occupied by strong magnetic fields.

\subsection{Sensitivity to Gaussian smoothing parameters}

The adopted temporal smoothing scale was examined by repeating the analysis with $\sigma_t$ values of 5, 20, and 27 days.
The overall latitude-dependent correlations, lag-correlation profiles, and cycle-to-cycle behavior remained qualitatively unchanged.
Increasing $\sigma_t$ systematically increased the absolute correlation coefficients, but did not alter the location of the active-latitude
correlation maxima or the broad positive-lag structure. Consequently, the main conclusions of the study are robust with respect
to the choice of temporal smoothing scale.

\subsection{Time-lag correlations}

The lag-correlation maxima and corresponding 95\% lag intervals are
listed in Table \ref{tab:lag95}.
Although formal lag maxima are found between 3 and 164 days,
the corresponding confidence intervals are generally broad and often
include zero lag.

\begin{table}
\flushleft
\caption{Lag-correlation maxima and the corresponding 95\% lag intervals for the Gaussian activity field ($\sigma_t = 11$ days).}
\label{tab:lag95}
\begin{tabular}{lccc}
\toprule
Cycle & Lag$_{\rm max}$ (days) & Lag$_{95,\min}$ (days) & Lag$_{95,\max}$ (days) \\
\midrule
C18 & 164 & -13 & 200 \\
C19 &  29 & -41 & 200 \\
C20 &  57 & -34 & 200 \\
C21 &  38 & -36 & 200 \\
C22 &   3 & -50 & 192 \\
C23 &   6 & -92 & 130 \\
C24 &  40 &  13 &  97 \\
\bottomrule
\end{tabular}
\end{table}

\begin{figure}[ht]
\centering
\includegraphics[width=1.0\textwidth]{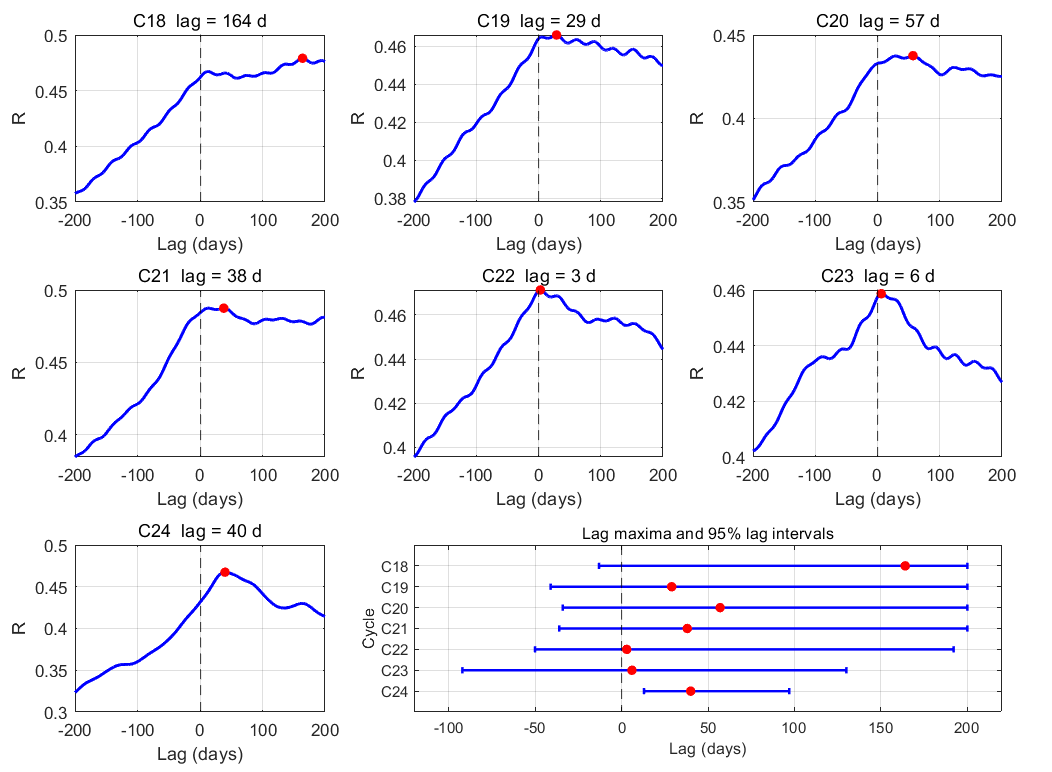}
\caption{Lag-correlation profiles between the green-line corona and the Gaussian sunspot activity field ($\sigma_t = 11$ d) for Solar Cycles 18--24. Red symbols indicate the formal correlation maxima. Although positive-lag maxima are present in most cycles, the profiles exhibit broad plateaus extending over tens of days rather than narrow isolated peaks. The corresponding 95 \% lag intervals are listed in Table~\ref{tab:lag95}. Several cycles also exhibit weak quasi-periodic fluctuations with a characteristic spacing comparable to one solar rotation ($\sim27$ days).}
\label{fig:lag}
\end{figure}

Figure~\ref{fig:lag}  presents the lag correlations between the green-line corona and the Gaussian sunspot activity field for cycles~18--24. All cycles exhibit a similar systematic behavior: the correlation increases toward positive lags and reaches a broad maximum approximately between 20 and 60 days with the exception of cycle 18.

Positive lag values imply that the coronal response follows sunspot activity. The broad plateau-like maximum indicates that the corona does not react instantaneously to the emergence of sunspot activity, but instead evolves on longer timescales comparable to one or more solar rotations.

The lag-correlation profiles are remarkably similar from cycle to cycle despite substantial differences in cycle amplitude. The highest correlations are obtained for cycles~21 and~24 (notice that this is incomplete), while cycle~20 exhibits systematically lower correlations over all lags.

No sharp isolated maximum is present in the universal lag profile. Instead, the correlation remains elevated over a broad range of positive lags. Correlation values near the solar rotation timescale ($\sim27$~days) are nearly as large as the formal maximum, suggesting that recurrent rotational modulation may contribute significantly to the observed lag structure.

\rev{\subsection{Statistical significance of lag maxima}}

Surrogate-data tests show that the formal lag maxima are generally not statistically distinct from the correlations obtained at zero lag. The lag-correlation curves therefore represent broad positive-lag plateaus rather than sharply defined delay times. Consequently, the lag maxima should be interpreted as indicators of temporal persistence of coronal activity rather than precise estimates of a physical delay between photospheric and coronal evolution.

\begin{figure}[ht]
\centering
\includegraphics[width=1.0\textwidth]{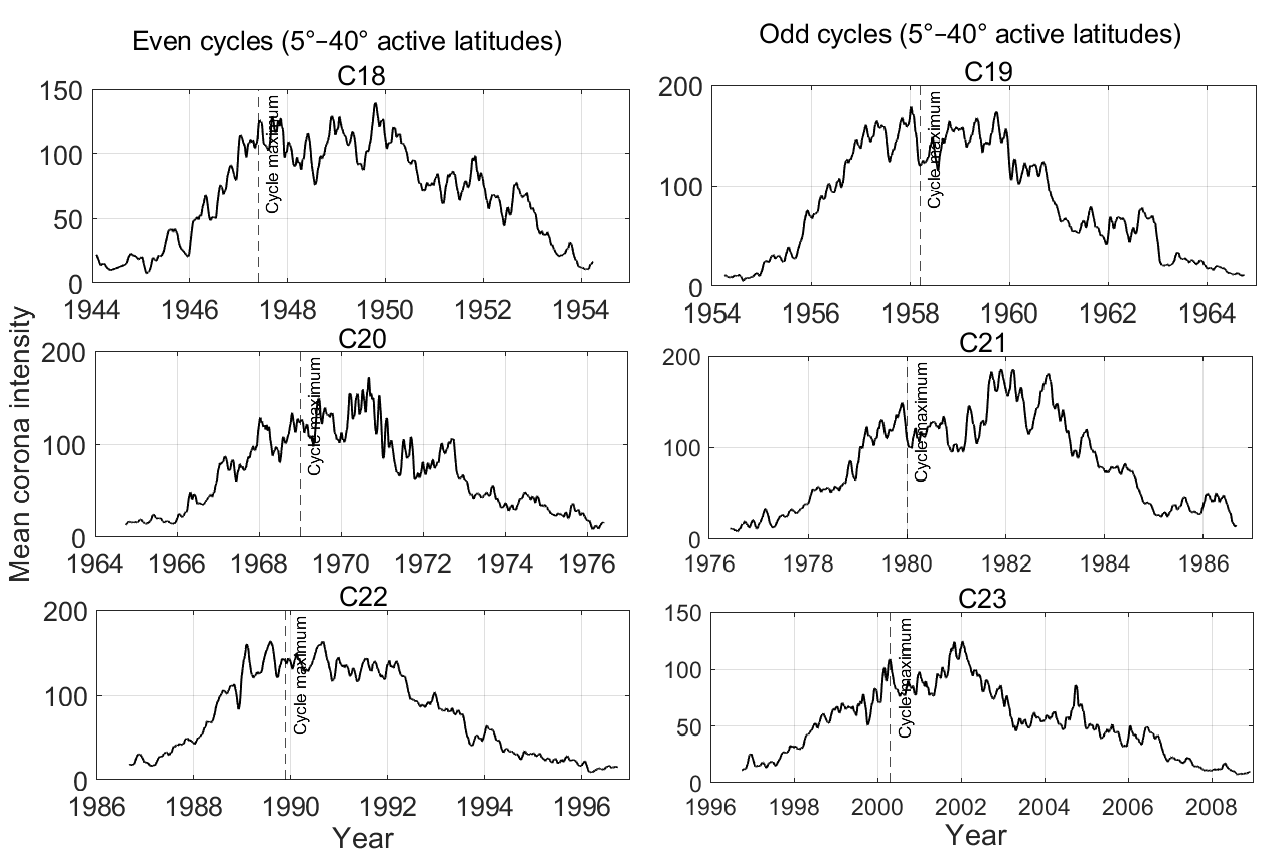}
\caption{Evolution of active-latitude ($5^\circ$--$40^\circ$) green-line coronal intensity during Solar Cycles 18--23. The vertical dashed lines indicate the epochs of maximum smoothed sunspot activity. In several cycles, the strongest coronal emission occurs months after the formal sunspot maximum, indicating that large-scale coronal magnetic structures persist beyond the peak phase of photospheric activity. The morphology of the maximum phase varies substantially from cycle to cycle, ranging from relatively distinct local reductions in coronal intensity to broad multi-peaked maxima. No unique coronal signature associated with the Gnevyshev-gap phenomenon is evident across all cycles.}
\label{fig:corona_max}
\end{figure}

\rev {\subsection{Coronal evolution during cycle maxima}}

Figure~\ref{fig:corona_max} illustrates the evolution of active-latitude ($5^\circ$--$40^\circ$) green-line coronal intensity during Solar Cycles~18--23. The vertical dashed lines indicate the epochs of maximum smoothed sunspot activity. In several cycles the strongest coronal emission does not coincide exactly with the sunspot maximum. Instead, the corona often remains enhanced for months after the formal sunspot maximum and exhibits complex multi-peaked structures during the maximum phase.

The behavior is particularly evident in Cycle~19, where coronal intensity shows a temporary reduction near the epoch of sunspot maximum before recovering to high values later in the cycle. Similar, although not identical, multi-peaked structures are visible in several other cycles. The timing, amplitude, and duration of these temporary reductions vary substantially from cycle to cycle. In some cases the maximum phase is characterized by a relatively distinct local minimum, whereas in others the corona forms a broader plateau with multiple intensity peaks.

These results indicate that the evolution of the green-line corona during solar maximum is more complex than that of traditional sunspot indices. The strongest coronal emission frequently occurs after the epoch of maximum sunspot activity, suggesting that large-scale coronal magnetic structures possess a longer temporal persistence than the photospheric activity responsible for sunspot production. This behavior is consistent with the positive lags obtained in the correlation analysis and supports the interpretation that the corona retains a significant magnetic memory of earlier photospheric activity.

The temporary reductions observed near cycle maximum may be related to the Gnevyshev-gap phenomenon. However, the comparison between different cycles demonstrates that no unique or universal coronal Gnevyshev-gap signature exists. Instead, the morphology of the maximum phase varies considerably from cycle to cycle, ranging from relatively sharp local reductions to broad multi-peaked maxima. Therefore, the present results suggest that the green-line corona reflects the general restructuring of the large-scale solar magnetic field during maximum conditions rather than a single well-defined manifestation of the Gnevyshev gap.

Surrogate tests were performed for representative cycles C19 and C24. In both cases the observed $\Delta R$ was not statistically distinguishable from surrogate realizations (p = 1.00 and p = 0.82), indicating that the maxima represent broad correlation plateaus rather than unique lag values.
\newline

\section{Discussion}

The present analysis demonstrates that the large-scale green-line corona is closely related to the spatiotemporal distribution of sunspot activity. The strongest correlations occur within the active-latitude belts, where the emergence of magnetic flux dominates the evolution of coronal structures. The Gaussian activity-field representation consistently produces stronger correlations than simple sunspot counts or sunspot areas, indicating that the corona responds primarily to spatially extended magnetic activity rather than to individual sunspots. This supports the interpretation that large-scale magnetic organization is a key factor governing coronal emission. A similar conclusion was reached by \citet{Badalyan2014}, who found that green-line coronal intensity is more closely related to large-scale magnetic fields than to local sunspot activity and that the strongest relationships occur within the active-latitude belts.

The latitude-dependent correlation profiles exhibit a stable double-peaked structure centered near the active-region belts. The similarity of this morphology across Solar Cycles 18--24 suggests that the latitude dependence of the corona--sunspot relationship is largely independent of cycle amplitude and parity. The strongest correlations occur near the core of the active latitudes, while weaker correlations are found both near the equator and at higher latitudes. This result is broadly consistent with the latitude distribution of strong photospheric magnetic flux reported by \citet{Vernova2016}, although the correlation profiles appear somewhat narrower than the corresponding magnetic-flux distributions. The concentration of the strongest correlations within the active-region belts is also in agreement with the findings of \citet{Badalyan2014}, who reported that the relationship between green-line emission and magnetic activity is strongest at low and mid-latitudes associated with active regions.

The relatively weak correlations near the equator are noteworthy because the green-line corona remains bright there during solar maximum. This apparent discrepancy suggests that strong coronal emission does not necessarily correspond to local sunspot activity. Instead, the equatorial corona likely reflects the combined influence of magnetic structures originating from both hemispheres together with long-lived streamers and large-scale magnetic connectivity. The reduced equatorial correlation therefore supports the view that the green-line corona behaves as a broadened large-scale magnetic response rather than a purely local tracer of sunspot emergence.

The lag-correlation analysis reveals broad positive-lag plateaus rather than narrow isolated maxima. Formal lag maxima are found in most cycles, but the corresponding 95 \% lag intervals are generally wide and often include zero lag. Furthermore, surrogate-data tests performed for representative cycles indicate that the observed lag maxima are not statistically distinguishable from neighbouring lag values. Consequently, the lag maxima should not be interpreted as precise physical delay times. Instead, the broad positive-lag plateaus are interpreted as evidence for temporal persistence and memory of large-scale coronal magnetic structures. This interpretation is consistent with the recurrent nature of coronal magnetic systems and their evolution on timescales comparable to one or more solar rotations. Several lag-correlation profiles exhibit weak quasi-periodic fluctuations with a characteristic spacing of approximately one solar rotation ($\sim27$ days). Although this behavior was not investigated further in the present work, it may reflect the rotational recurrence and persistence of large-scale coronal magnetic structures. Such recurrence is consistent with the long-lived nature of active-region complexes and coronal streamer systems, which can remain organized over multiple solar rotations.

The analysis of active-latitude coronal intensity during cycle maxima demonstrates that the strongest coronal emission often occurs after the formal sunspot maximum. This behavior is particularly evident in Cycle~19 but is also visible in several other cycles. The morphology of the maximum phase varies considerably from cycle to cycle, ranging from relatively distinct local reductions to broad multi-peaked maxima. These results suggest that no unique coronal signature of the Gnevyshev gap exists. Instead, the observed variations appear to reflect cycle-dependent reorganization of the large-scale solar magnetic field during maximum conditions.

An additional indication of coronal persistence is provided by the relationship between the lag-correlation maxima and the timing of the coronal cycle maxima. Cycles exhibiting larger delays between the sunspot maximum and the subsequent coronal maximum also tend to show larger positive lag-correlation maxima. Although the number of available cycles is small and the relationship is influenced by outliers such as Cycle 18, this tendency suggests that both quantities may reflect a common underlying property, namely the temporal persistence of large-scale coronal magnetic structures.

Solar Cycle~24 differs somewhat from the earlier cycles. Although the overall latitude dependence remains similar, the lag-correlation profile reaches its maximum closer to zero lag and declines more rapidly at positive lags. This behavior may indicate a shorter temporal persistence of large-scale coronal structures during the unusually weak Cycle~24, implying a reduced coronal memory compared with stronger cycles.

Overall, the results suggest that the green-line corona can be quantitatively described using continuous latitude--time representations of sunspot activity. The persistence of the latitude-dependent and lag-dependent structures across multiple solar cycles indicates that these relationships reflect fundamental properties of the global solar magnetic cycle rather than cycle-specific phenomena. The results are also consistent with recent studies by \citet{Oloketuyi2024} and \citet{Oloketuyi2025}, which reported latitude-dependent relationships and temporal persistence between coronal and photospheric manifestations of solar activity. Together with the earlier results of \citet{Badalyan2014} and \citet{Vernova2016}, the present analysis supports the view that the green-line corona represents a large-scale magnetic response whose evolution is governed by the spatial organization and long-term persistence of solar magnetic activity.

The latitude dependence is broadly consistent with the latitude distribution of strong photospheric magnetic flux reported by Vernova et al. (2016), although the present correlation profile is more narrowly concentrated within the core active-latitude belts.

\section{Conclusions}

The results demonstrate that the green-line corona is not simply a local tracer of individual sunspots, but instead behaves as a broadened large-scale response to the spatial distribution of magnetic activity. The strongest correlations between the corona and sunspot activity are concentrated within the active-region belts, while weaker but significant correlations extend toward higher latitudes. Compared with traditional sunspot-number or sunspot-area representations, the Gaussian activity-field model produces systematically stronger and more coherent latitude--time correlations, indicating that the large-scale organization of magnetic activity is an essential component of coronal structure formation.

The latitude-dependent correlation profile exhibits a stable double-peaked morphology centered on the active-latitude belts. This structure remains remarkably similar across Solar Cycles 18--24 and is broadly consistent with the latitude distribution of strong photospheric magnetic flux reported by Vernova et al. (2016), although the correlation peaks are somewhat more localized in latitude.

The lag-correlation profiles exhibit broad positive-lag plateaus in most cycles. Surrogate-data tests indicate that the formal lag maxima are generally not statistically distinguishable from neighboring lag values and should therefore not be interpreted as precise physical delays. Instead, the observed lag structure is interpreted as evidence for temporal persistence and memory of large-scale coronal magnetic systems. The broad lag-correlation plateaus and wide 95 \% lag intervals indicate that the observed positive lags should be interpreted as evidence of coronal persistence rather than as precisely defined physical delay times.

The active-latitude coronal intensity often reaches its strongest values after the formal sunspot maximum. This behavior is particularly evident in Cycle 18 and several subsequent cycles, suggesting that large-scale coronal structures may persist beyond the peak phase of photospheric activity. The morphology of cycle maxima varies substantially from cycle to cycle, however, indicating that no unique coronal signature of the Gnevyshev-gap phenomenon exists.

Overall, the results support the interpretation that the green-line corona represents a temporally persistent large-scale magnetic system whose evolution is governed by the distributed emergence, reorganization, and long-term evolution of solar magnetic activity.

\section*{Acknowledgements}

The author thanks the data providers for making the corona and sunspot data available. No external funding was received for this study.

\bibliographystyle{plainnat}
\bibliography{references}

@article{Badalyan2008,
author = {Badalyan, O. G. and Obridko, V. N. and Sykora, J.},
title = {Quasibiennial Oscillations in the North--South Asymmetry of Solar Activity},
journal = {Solar Physics},
year = {2008},
volume = {247},
pages = {379--397},
doi = {10.1007/s11207-007-9103-3}
}

@article{Badalyan2005,
  author = {Badalyan, O. G. and Obridko, V. N. and Rybak, J. and Sykora, J.},
  title = {Quasibiennial Oscillations of the North--South Asymmetry},
  journal = {Astronomy Reports},
  year = {2005},
  volume = {49},
  pages = {659--670},
  doi = {10.1134/1.2038678}
}

@article{Badalyan2014,
  author  = {Badalyan, O. G. and Bludova, N. G.},
  title   = {Relation of the Green Coronal Line Intensity to Sunspot Areas and Magnetic Fields of Different Scales},
  journal = {Solar System Research},
  year    = {2014},
  volume   = {48},
  number   = {4},
  pages    = {305--315},
  doi      = {10.1134/S0038094614040029}
}

@article{Takalo2022,
  author  = {Takalo, J.},
  title   = {Spatial and Temporal Distribution of Solar Green-Line Corona for Solar Cycles 18--24},
  journal = {Solar Physics},
  year    = {2022},
  volume  = {297},
  number  = {118},
  doi     = {10.1007/s11207-022-02050-0}
}

@article{Rusin2002,
  author  = {Ru{\v s}in, V. and Rybansk{\'y}, M.},
  title   = {The Green Corona and Magnetic Fields},
  journal = {Solar Physics},
  year    = {2002},
  volume  = {207},
  pages   = {47--71},
  doi     = {10.1023/A:1015552009286}
}

@article{Rybansky2001,
  author  = {Rybansk{\'y}, M. and Ru{\v s}in, V. and Minarovjech, M.},
  title   = {Coronal Index of Solar Activity -- Solar-Terrestrial Research},
  journal = {Space Science Reviews},
  year    = {2001},
  volume  = {95},
  pages   = {227--234},
  doi     = {10.1023/A:1005238531081}
}

@article{Vernova2016,
  author  = {Vernova, E. S. and Tyasto, M. I. and Baranov, D. G.},
  title   = {Latitudinal Distribution of Photospheric Magnetic Fields of Different Magnitudes},
  journal = {Solar Physics},
  year    = {2016},
  volume  = {291},
  pages   = {741--756},
  doi     = {10.1007/s11207-016-0863-1}
}

@article{Kane2015,
  author  = {Kane, R. P.},
  title   = {Solar Cycle Variation of Coronal Green Line Index},
  journal = {Indian Journal of Radio and Space Physics},
  year    = {2015},
  volume  = {44},
  pages   = {122--128}
}

@article{Gnevyshev1977,
  author  = {Gnevyshev, M. N.},
  title   = {Essential Features of the 11-Year Solar Cycle},
  journal = {Solar Physics},
  year    = {1977},
  volume  = {51},
  pages   = {175--183},
  doi     = {10.1007/BF00240455}
}

@article{Feminella1997,
  author  = {Feminella, F. and Storini, M.},
  title   = {Large-Scale Dynamical Phenomena During Solar Activity Cycles},
  journal = {Astronomy and Astrophysics},
  year    = {1997},
  volume  = {322},
  pages   = {311--319}
}

@article{Ahluwalia2004,
  author  = {Ahluwalia, H. S. and Kamide, Y.},
  title   = {Gnevyshev Gap, Forbush Decreases, ICMEs and Solar Wind Electric Field: Relationships},
  journal = {Advances in Space Research},
  year    = {2004},
  volume  = {35},
  pages   = {2119--2124},
  doi     = {10.1016/j.asr.2004.12.005}
}

@article{TakaloMursula2020,
  author  = {Takalo, J. and Mursula, K.},
  title   = {Comparison of the Shape and Temporal Evolution of Even and Odd Solar Cycles},
  journal = {Astronomy and Astrophysics},
  year    = {2020},
  volume  = {636},
  pages   = {A11},
  doi     = {10.1051/0004-6361/201937124}
}

@article{Minarovjech2011,
  author  = {Minarovjech, M. and Ru{\v s}in, V. and Saniga, M.},
  title   = {The Green Corona Database and the Coronal Index of Solar Activity},
  journal = {Contributions of the Astronomical Observatory Skalnat{\'e} Pleso},
  year    = {2011},
  volume  = {28},
  pages   = {137--148}
}

@ARTICLE{Lukac_2010,
       author = {{Luk{\'a}{\v{c}}}, B. and {Rybansk{\'y}}, M.},
        title = "{Modified Coronal Index of the Solar Activity}",
      journal = SolPhys,
         year = 2010,
        month = may,
       volume = {263},
       number = {1-2},
        pages = {43-49},
          doi = {10.1007/s11207-010-9545-0},
       adsurl = {https://ui.adsabs.harvard.edu/abs/2010SoPh..263...43L}
}

@ARTICLE{Rybansky2005,
       author = {{Rybansk{\'y}}, M. and {Ru{\v{s}}in}, V. and {Minarovjech}, M. and {Klocok}, L. and {Cliver}, E.~W.},
        title = "{Reexamination of the coronal index of solar activity}",
      journal = {Journal of Geophysical Research (Space Physics)},
         year = 2005,
        month = aug,
       volume = {110},
       number = {A8},
          eid = {A08106},
        pages = {A08106},
          doi = {10.1029/2005JA011146},
       adsurl = {https://ui.adsabs.harvard.edu/abs/2005JGRA..110.8106R}
}

@article{Oloketuyi2025,
  author = {Oloketuyi, J. and Liu, Y. and Deng, L. and Elmhamdi, A. and Almosabeh, K. N.r and Zhu, F.g and Li, H. and Sha, F. and Liu, Q. and Abe, O. E. and Owolabi, C. and Olusola, O.},
  title = {Temporal Behavior and Latitudinal Relationships between Key Solar Parameters and Green Line Emissions in the Solar Corona},
  journal = {The Astrophysical Journal Letters},
  year = {2025},
  volume = {984},
  number = {1},
  pages = {L34},
  doi = {10.3847/2041-8213/adcb38}
}

@article{Oloketuyi2024,
  author = {Oloketuyi, J. and Liu, Y. and Elmhamdi, A. and Zhu, F. and Deng, L.},
  title = {Understanding the long-term evolution of green line coronal emission and its relation to the sunspots},
  journal = {Astrophysics and Space Science},
  year = {2024},
  volume = {369},
  number = {4},
  pages = {35},
  doi = {10.1007/s10509-024-04300-y}
}

@article{Morgan2017,
  author  = {Morgan, H. and Taroyan, Y.},
  title   = {Global Conditions in the Solar Corona from 2010 to 2017},
  journal = {Science Advances},
  year    = {2017},
  volume   = {3},
  number   = {7},
  pages    = {e1602056},
  doi      = {10.1126/sciadv.1602056}
}

@article{Hathaway2015,
  author  = {Hathaway, David H.},
  title   = {The Solar Cycle},
  journal = {Living Reviews in Solar Physics},
  year    = {2015},
  volume   = {12},
  number   = {1},
  pages    = {4},
  doi      = {10.1007/lrsp-2015-4}
}

@book{Silverman1986,
author = {Silverman, B. W.},
title = {Density Estimation for Statistics and Data Analysis},
publisher = {Chapman and Hall},
address = {London},
year = {1986},
isbn = {0-412-24620-1}
}

\end{document}